\documentclass[conference]{IEEEtran} 
\pdfoutput=1

\usepackage[T1]{fontenc}
\usepackage[utf8]{inputenc}

\usepackage{amsmath,amssymb,amsfonts}
\usepackage[export]{adjustbox}
\usepackage{booktabs}
\usepackage{cite}
\usepackage{enumitem}
\usepackage{graphicx}
\usepackage{placeins}
\usepackage{tabularx}
\usepackage{url}
\usepackage[dvipsnames]{xcolor}

\usepackage{efbox}
\efboxsetup{linecolor=black,linewidth=0.5pt}

\usepackage{glossaries}
\glsdisablehyper
\newacronym{1ch}{\mbox{1-CH}}{single-channel}
\newacronym{1pps}{\mbox{1 PPS}}{1 pulse per second}
\newacronym{a2g}{\mbox{A2G}}{air-to-ground}
\newacronym{a2a}{\mbox{A2A}}{air-to-air}
\newacronym{b2b}{\mbox{B2B}}{Back-to-back}
\newacronym{cir}{\mbox{CIR}}{channel impulse responses}
\newacronym{cots}{\mbox{COTS}}{commercial off-the-shelf}
\newacronym{doa}{\mbox{DoA}}{direction-of-arrival}
\newacronym{fpga}{\mbox{FPGA}}{field programmable gate array}
\newacronym{gnssdo}{\mbox{GNSSDO}}{GNSS disciplined Oscillator}
\newacronym{ins}{\mbox{INS}}{inertial navigation system}
\newacronym{isac}{\mbox{ISAC}}{integrated sensing and communication}
\newacronym{los}{\mbox{LoS}}{line-of-sight}
\newacronym{lna}{\mbox{LNA}}{low noise amplifier}
\newacronym{mimo}{\mbox{MIMO}}{multiple-input multiple-output}
\newacronym{mqtt}{\mbox{MQTT}}{message queueing telemetry transport}
\newacronym{siso}{\mbox{SISO}}{single-input single-output}
\newacronym[plural=\mbox{SDRs},firstplural=software defined \mbox{radios (SDR)}]{sdr}{\mbox{SDR}}{software defined radio}
\newacronym{ssd}{\mbox{SSD}}{solid-state disk}
\newacronym{snr}{\mbox{SNR}}{signal-to-noise ratio}
\newacronym{rtk}{\mbox{RTK}}{real-time kinematic}
\newacronym{rpc}{\mbox{RPC}}{remote procedure call}
\newacronym{rf}{\mbox{RF}}{radio frequency}
\newacronym{rx}{\mbox{Rx}}{receiver}
\newacronym{tdma}{\mbox{TDMA}}{time division multiple access}
\newacronym{toa}{\mbox{ToA}}{time-of-arrival}
\newacronym{tx}{\mbox{Tx}}{transmitter}
\newacronym[plural=\mbox{UAVs}]{uav}{\mbox{UAV}}{unmanned aerial vehicle}
\newacronym{ue}{\mbox{UE}}{user equipment}
\newacronym{usrp}{\mbox{USRP}}{Universal Software Radio Peripheral}
\newacronym{v2x}{\mbox{V2X}}{vehicle-to-everything} 
\newacronym{vtol}{\mbox{VTOL}}{vertical take-off and landing}
\newacronym{vrus}{\mbox{VRUs}}{vulnerable road users} 

\usepackage{hyperref}
\hypersetup{
 	pdfauthor={Julia Beuster},
 	pdftitle={ },
     pdfkeywords={Integrated sensing and communication (ISAC), 6G, distributed multi-sensor network, propagation measurements, channel sounding, unmanned aerial vehicle, air-to-air (A2A), air-to-ground (A2G), radar sensing, software defined radio (SDR), USRP, real-time kinematic (RTK)}
 }

\usepackage{orcidlink}

\title{Enhancing Situational Awareness in ISAC Networks via Drone Swarms: A Real-World Channel Sounding Data Set}

\IEEEoverridecommandlockouts

\author{%
    \IEEEauthorblockN{%
        Julia~Beuster\raisebox{.5ex}{\orcidlink{0000-0003-1887-4278}}\IEEEauthorrefmark{1},
        Carsten~Andrich\raisebox{.5ex}{\orcidlink{0000-0002-4795-3517}}\IEEEauthorrefmark{1},
        Sebastian~Giehl\raisebox{.5ex}{\orcidlink{0009-0008-1672-1351}}\IEEEauthorrefmark{1},
        Marc~Miranda\raisebox{.5ex}{\orcidlink{0000-0000-0000-0000}}\IEEEauthorrefmark{1}, 
        Lorenz~Mohr\raisebox{.5ex}{\orcidlink{0009-0008-2599-6891}}\IEEEauthorrefmark{1}, \\
        Dieter~Novotny\raisebox{.5ex}{\IEEEauthorrefmark{2}},
        Tom~Kaufmann\raisebox{.5ex}{\IEEEauthorrefmark{3}},
        Christian~Schneider\raisebox{.5ex}{\orcidlink{0000-0003-1833-4562}}\IEEEauthorrefmark{1},
        Reiner~S.~Thomä\raisebox{.5ex}{\orcidlink{0000-0002-9254-814X}}\IEEEauthorrefmark{1}
    }
    \IEEEauthorblockA{\IEEEauthorrefmark{1}Technische Universität Ilmenau, Institute for Information Technology, Ilmenau, Germany, \\ 
    \IEEEauthorrefmark{2}AeroDCS Gmbh, Koblenz, Germany, 
    \IEEEauthorrefmark{3}CiS GmbH, Rostock, Germany
    }    
    \thanks{This research has been partially funded by the Federal Ministry of Education and Research of Germany in the projects ``6G-ICAS4Mobility'' (grant number: 16KISK241).}
}

\begin{document}
\maketitle

\begin{abstract}
With the upcoming capabilities of \gls{isac} and the incorporation of \mbox{\gls{ue}} like \glspl{uav} in 6G~mobile networks, there is a significant opportunity to enhance situational awareness through multi-static radar sensing in meshed \gls{isac} networks.
This paper presents a real-world channel sounding data set acquired using a testbed with synchronized, distributed ground-based sensor nodes and flying sensor nodes within a swarm of up to four drones.
The conducted measurement campaign is designed to sense the bi-static reflectivity of objects such as parking cars, \mbox{\gls{vtol}} aircraft, and small drones in multi-path environments.
We detail the rationale behind the selection of the included scenarios and the configuration of the participating nodesand present exemplary results to demonstrate the potential of using collaborating drone swarms for multi-static radar tracking and localization in \gls{a2a} and \gls{a2g} scenarios.
The data sets are publicly available to support the development and validation of future \gls{isac} algorithms in real-world environments rather than relying solely on simulation.
\end{abstract}


\section{Introduction}
\begin{figure}[b!]
\centering
	\includegraphics[width=0.89\linewidth] {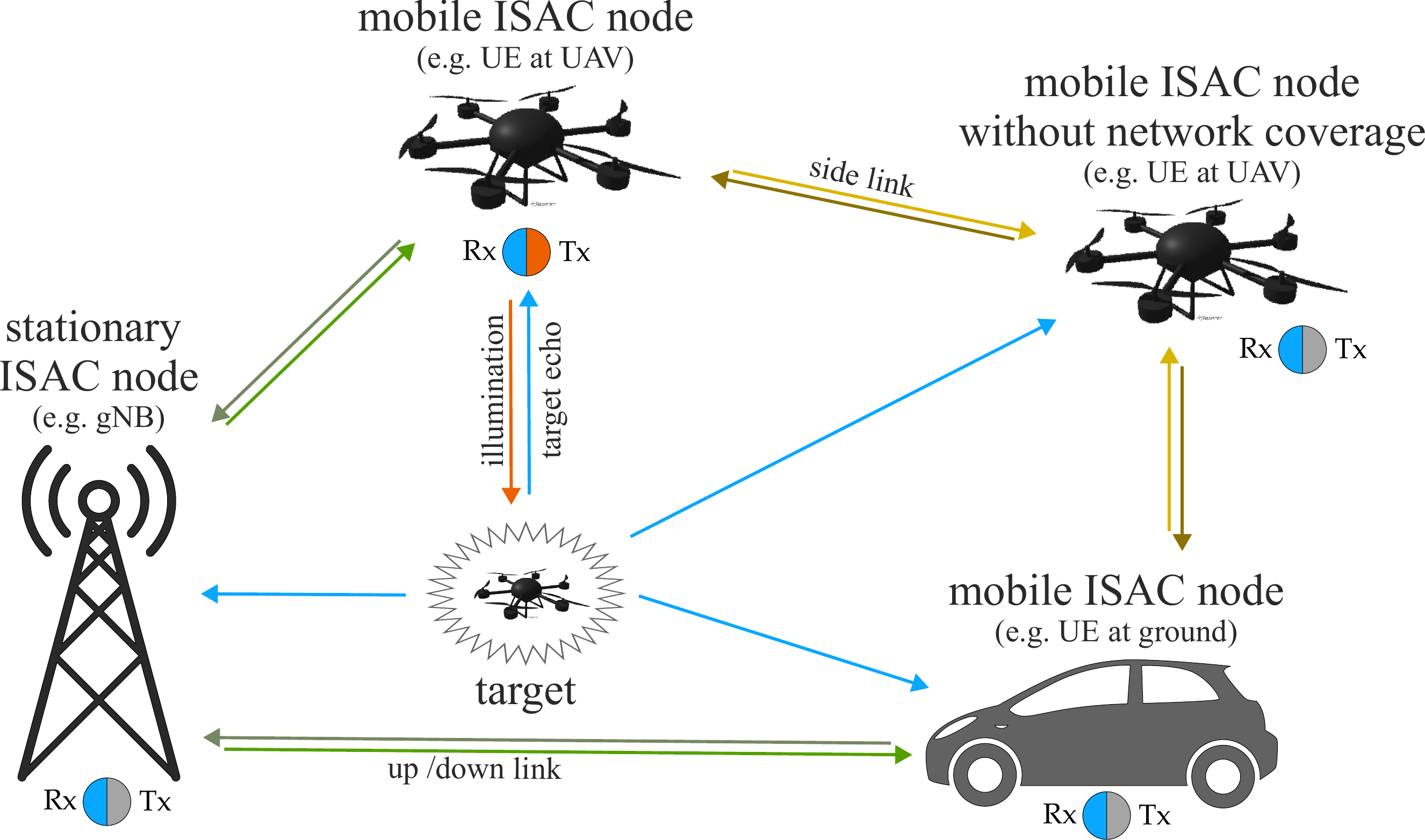}
	\caption{Illustration of a multi-sensor ISAC network featuring stationary and mobile transceiver nodes for  infrastructure-based and mobile node sensing. The Tx node illuminates a target to allow radar sensing at the Rx nodes. Typically, all participating nodes can operate as Rx and Tx.
    }
\label{isacnetwork}
\end{figure}

The exploitation of \gls{isac} functionalities in the mobile radio landscape offers a promising approach to enhance situational awareness across various applications, such as traffic monitoring, search and rescue operations, public safety, and aerial surveillance, while optimizing the use of scarce resources~\cite{10646287}.
In particular, the integration of mobile airborne nodes offers a promising approach to detect and localize passive objects through their bi-static back-scattering and enable multi-static radar sensing in meshed \gls{isac} networks, as shown in Fig.~\ref{isacnetwork}.
This is especially interesting for detecting challenging objects with low radar cross-section and high mobility such as small UAVs~\cite{10938573}.
The availability of realistic and diverse data sets capturing the complexities of real-world scenarios in combination with data from channel modeling and simulations is essential for the development and validation of \gls{isac} algorithms prior to live system integration to avoid the cost and repetition of extensive measurement campaigns~\cite{10107507}.
Recent studies have introduced several channel sounding data sets featuring small airborne nodes at \glspl{uav} in \gls{a2g} scenarios and \gls{a2a} scenarios with large airborne nodes like helicopters~\cite{10616106}.
Unfortunately, these measurement campaigns were not tailored for radar sensing in \gls{isac} applications and therefore lack passive objects with reliable positioning data.

This paper presents \gls{a2a} sounding data sets that capture the dynamics of sensing in a \gls{uav} swarm consisting of up to four transport hexacopters flying in various constellations and \gls{a2g} measurement data sets for infrastructure-based sensing using additional stationary sensor nodes at exposed positions at rooftops and on the ground.
A synchronized channel sounding testbed with distributed sensor nodes, introduced in~\cite{beuster2025realtimesoundingisacnetworks}, was used to collect data for sensing dynamic and static airborne objects such as \gls{vtol} aircraft, and a small quadcopter, as well as ground-based objects such as cars, in a multi-path environment.
We outline the steps and configurations needed to plan and execute a real-world measurement campaign and present exemplary results using the acquired data set.
The data is publicly available at~\cite{EMS_dataset}.

\section{Measurement Campaign Design}
To ensure the acquisition of reliable test data for \gls{isac} networks with airborne nodes, it is essential to select, configure and calibrate the measurement system, as well as strategically plan the various flight trajectories and environmental conditions during the real-world measurement.

\subsection{Sounder Testbed Configuration}
\begin{figure}[t]
	\includegraphics[width=\linewidth] {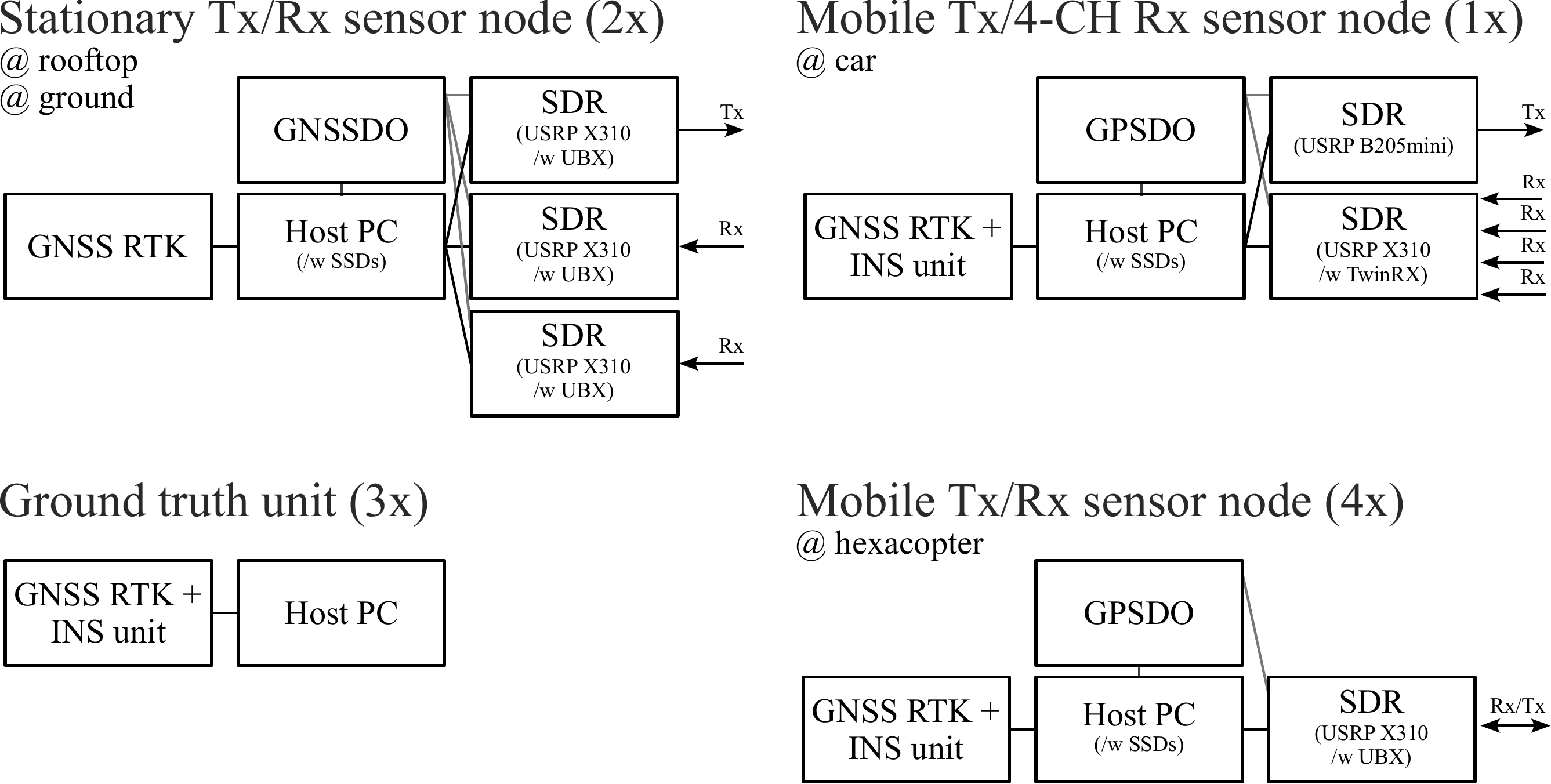}
	\caption{Simplified block diagrams of a multi-node sounding testbed with distributed synchronized transceiver nodes and highly accurate ground truth for all participating nodes, as introduced in \cite{beuster2025realtimesoundingisacnetworks}. The number of nodes were mainly limited by available hardware and staff resources.
    }
\label{setup}
\end{figure}

\noindent A sounding testbed for \gls{isac} applications needs to ensure reliable continuous real-time recording of \gls{cir} with large bandwidths, consist of a modular system of synchronized transceiver nodes with configurable Rx and Tx signal paths offering flexibility in number of nodes and compatibility with various carrier systems, and record highly accurate positioning and orientation data for reliable ground truth data for all participating nodes.
The measurement testbed introduced in~\cite{beuster2025realtimesoundingisacnetworks} offers a modular and scalable system of GNSS synchronized \mbox{\gls{sdr}-based} stationary and mobile transceiver nodes with centimeter-level \gls{rtk} and \gls{ins}-based positioning accuracy, as well as a \gls{rpc} interface for all participating nodes.
The mobile nodes range from light-weight to medium-weight to allow the deployment on different carrier systems, including \glspl{uav}.
The measurement campaign involved two stationary multi-channel sensor nodes mounted on tripods at the rooftop of a building and on the ground at an adjacent parking lot, four mobile single-channel sensor nodes mounted at medium-sized transport hexacopters, one mobile multi-channel sensor node mounted at a car and three ground truth units for the participating targets, as illustrated in Fig.~\ref{setup}. 
The participating sensor nodes were equipped with different types of antennas and the overall sensor node placement and signal parameter configuration were chosen to provide conditions and signal parameters similar to those present for 5G~\gls{ue} in FR1, as summarized in Table~\ref{tab:specs_meas}.
The antenna patterns were measured in an anechoic chamber and \gls{b2b} calibration performed for all Tx/Rx combinations.
\begin{table}[t]
	\caption{Summary of measurement campaign characteristics.}
	\vspace{-1em}
	\centering
    \begin{tabularx}{\linewidth}{Xl}
	\toprule
    \small Measurement site & \\
    \multicolumn{2}{l}{\quad -- 2-story building at $\sim$100$\times$100m parking lot}\\
    \midrule
 	\small Signal parameters & OFDM signal \scriptsize(Newman phase sequence) \\
 	\quad -- Center frequency & 3.75\,GHz \\
    \quad -- Symbol length & 16 \textmu s \\
 	\quad -- Bandwidth (used) & 80 MHz, 48 MHz \\
    \quad -- Subcarriers & 1280, 768 \\
	\quad -- EIRP & 46 dBm, 23 dBm \\
    \midrule
    \multicolumn{2}{l}{\small Stationary sensor nodes\footnotemark[1]}\\
    \quad -- Tx\footnotemark[2] \mbox{({\color{BrickRed} $\blacktriangle$})} at rooftop & \scriptsize antenna: 23\textdegree~uptilt, 11.5\,m (AGL)\\
    \quad -- Rx\footnotemark[2] \mbox{({\color{BurntOrange} $\blacktriangle$})} \\
    \qquad \quad \#01, \#02 at rooftop & \scriptsize antenna:  25\textdegree~uptilt, 9.9\,m (AGL)\\
    \qquad \quad \#03, \#04 at ground & \scriptsize antenna: 40\textdegree and 41\textdegree~uptilt, 1.9 and 2.4\,m (AGL)\\
    \midrule
    \multicolumn{2}{l}{\small Mobile sensor nodes\footnotemark[1]}\\
    \multicolumn{2}{l}{\quad -- Tx\footnotemark[3] \mbox{({\color{MidnightBlue} $\bullet$})} at hexacopter}\\
    \multicolumn{2}{l}{\quad -- Rx\footnotemark[3]\mbox{({\color{Green} $\bullet$})} \#01, \#02, \#03  at hexacopter}\\
    \multicolumn{2}{l}{\quad -- 4-CH Rx\footnotemark[4] \mbox{({\color{LimeGreen} $\bullet$})} at car}\\
    \midrule
    \multicolumn{2}{l}{\small Targets \scriptsize (with GNSS ground truth unit) }\\
    \multicolumn{2}{l}{\quad -- VTOL \mbox{({\tiny \color{Purple} $\blacksquare$})} in the air}\\
    \multicolumn{2}{l}{\quad -- small quadcopter \mbox{({\tiny \color{RedViolet} $\blacksquare$})} {\scriptsize(DJI Matrice)} in the air}\\
    \multicolumn{2}{l}{\quad -- mid-sized car \mbox{({\tiny \color{RubineRed} $\blacksquare$})} in parking lot}\\
    \bottomrule
    \multicolumn{2}{l}{\footnotemark[1] \scriptsize Back-to-back (B2B) calibration between all Rx/Tx combinations}\\
	\multicolumn{2}{l}{\footnotemark[2] \scriptsize With tiltable, directional antenna (10~dB beamwidth at 40 degree)}\\
    \multicolumn{2}{l}{\footnotemark[3] \scriptsize With omni-directional antenna}\\
    \multicolumn{2}{l}{\footnotemark[4] \scriptsize With antenna array}\\
 	\end{tabularx}
 \label{tab:specs_meas}
 \end{table}
\subsection{Scenario Planning}
\begin{figure*}[h!]
	\includegraphics[width=\linewidth] {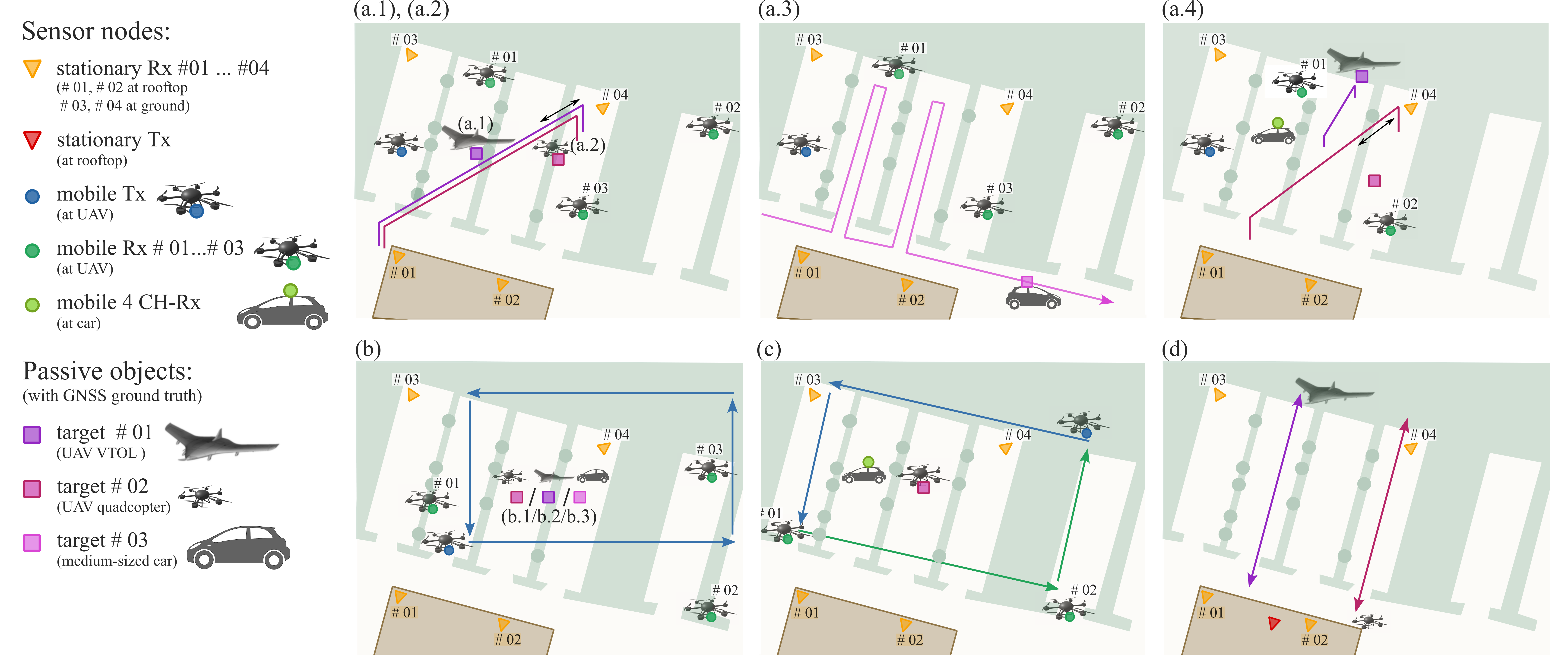}
	\caption{Simplified illustration of the measurement scenarios included in the data set (see Table~\ref{tab:scenarios}).
    }
\label{scenario}
\end{figure*}

The measurement campaign included several scenarios, each designed to capture different aspects of \gls{a2a} and \gls{a2g} sensing of ground-based and airborne targets, as summarized in Table~\ref{tab:scenarios}.
As shown in Fig.~\ref{scenario}, the flight trajectories of the collaborating drone swarm equipped with sensor nodes were selected to cover fixed-point hovering, straight-line flight, circular flight, and complex flight maneuvers.
These trajectories varied in terms of altitude, speed, and direction to diversify the data set by offering multiple angular perspectives, which is essential for the research of multi-static radar tracking and mitigation of blind spots.
The involved targets were chosen to be a common mid-sized car, a popular commercially available camera quadcopter and a \gls{vtol} aircraft, recently featured in a study about of bi-static Micro-Doppler analysis in \gls{isac} frameworks~\cite{10979139}, and their trajectories were designed to capture the propagation characteristics of passive objects moving parallel and perpendicular in close proximity to the sensors at varying altitudes and velocities during realistic maneuvers, for instance in take-off and landing situations in departure and arrival scenarios.
\begin{table*}[t]
	\caption{Overview of measurement scenarios in the data set}
    \vspace{-1em}
	\centering
    \begin{tabularx}{\linewidth}{lccccccccX}
    \toprule
    \small Scenario & \multicolumn{2}{l}{\small Stationary} & \multicolumn{3}{l}{\small Mobile} &\multicolumn{3}{l}{\small Targets} & \small Description\\
     & Tx & Rx & Tx & Rx & 4-CH Rx & VTOL & quadcopter & car & \scriptsize * all altitudes in meter above ground level (AGL)\\
     & \mbox{({\color{BrickRed} $\blacktriangle$})} & \mbox{({\color{BurntOrange} $\blacktriangle$})} & \mbox{({\color{MidnightBlue} $\bullet$})} & \mbox{({\color{Green} $\bullet$})} & \mbox{({\color{LimeGreen} $\bullet$})} & \mbox{({\tiny \color{Purple} $\blacksquare$})} & \mbox{({\tiny \color{RedViolet} $\blacksquare$})} & \mbox{({\tiny \color{RubineRed} $\blacksquare$})} & \\
     \midrule
     \multicolumn{9}{l}{\small (a) Mobile sensing of dynamic targets using a static UAV swarm} & Dynamic targets cross the field-of-view of a stationary UAV swarm. The UAVs with sensor nodes are hovering at 37..42\,m altitude.\\
     \par\vspace{-0.7em}\\
     \qquad (a.1) & - & 4 & 1 & 3 & - & 1 & - & - & Flying target at 27\,m incl. take-off and landing \\
     \qquad (a.2) & - & 4 & 1 & 3 & - & - & 1 & - & Flying target at 28\,m incl. take-off and landing\\
     \qquad (a.3) & - & 4 & 1 & 3 & - & - & - & 1 & Driving target in parking lot at ground level\\
     \qquad (a.4) & - & 4 & 1 & 2 & 1 & 1 & 1 & - & Multi-target scenario with flying VTOL at 45\,m and flying quadcopter at 27\,m, both incl. take-off and landing. The mobile multi-channel sensor node is mounted at a parked car.\\
     \midrule
     \multicolumn{9}{l}{\small (b) Mobile sensing of static targets using a partially dynamic UAV swarm} & Stationary targets are sensed with Rx nodes at hovering UAVs at 26..28\,m using the illumination by the mobile Tx mounted on a UAV flying at 37..40\,m.\\
     \par\vspace{-0.7em}\\
     \qquad (b.1) & - & 4 & 1 & 3 & - & - & 1 & - & Target hovering at 28\,m.\\
     \qquad (b.2) & - & 4 & 1 & 3 & - & 1 & - & - & Target hovering at 27\,m.\\
     \qquad (b.3) & - & 4 & 1 & 3 & - & - & - & 1 & Target parked in parking lot at ground level.\\
     \midrule
     \multicolumn{9}{l}{\small (c) Mobile sensing of static targets using a dynamic UAV swarm} & Stationary targets are sensed with mobile sensor nodes flying around the target at 31..43\,m altitude and a multi-channel sensor node mounted at a parked car at the parking lot.\\
     \par\vspace{-0.7em}\\
     \qquad (c.1) & - & 4 & 1 & 2 & 1 & - & 1 & - & Target hovering at 28\,m.\\
     \midrule
     \multicolumn{9}{l}{\small (d) Infrastructure-based sensing of flying targets} & Dynamic targets are sensed using the stationary sensor nodes at the rooftop and in the parking lot.\\
     \par\vspace{-0.7em}\\
     \qquad (d.1) & 1 & 4 & - & - & - & 1 & 1 & - & Multi-target scenario with flying VTOL at 31\,m and flying quadcopter at 27\,m in perpendicular flight during arrival/departure.\\
    \bottomrule
 	\end{tabularx}
 \label{tab:scenarios}
 \end{table*}
\section{Multi-static radar tracking of flying targets}
The recorded data set \cite{EMS_dataset} comprises 
\begin{itemize}
    \item[--] data files with pre-processed measurement data that considers calibration and synchronization data, and 
    \item[--] metadata files with extensive details about node configurations, trajectories and environmental conditions 
\end{itemize}
that allow to investigate the potential of multi-static sensing in \gls{isac} networks in complex environments, as shown for illustrative results from exemplarily chosen scenarios in Fig.~\ref{fig:results_a1} and Fig.~\ref{fig:results_d1}.
Considering the wide range of radar-based localization techniques, assessing individual algorithms with the measurement data is not within the scope of this paper.

\subsection{Mobile sensing in a drone swarm -- Scenario (a.1)}
\begin{figure*}[h]
    \includegraphics[width=0.3\linewidth]{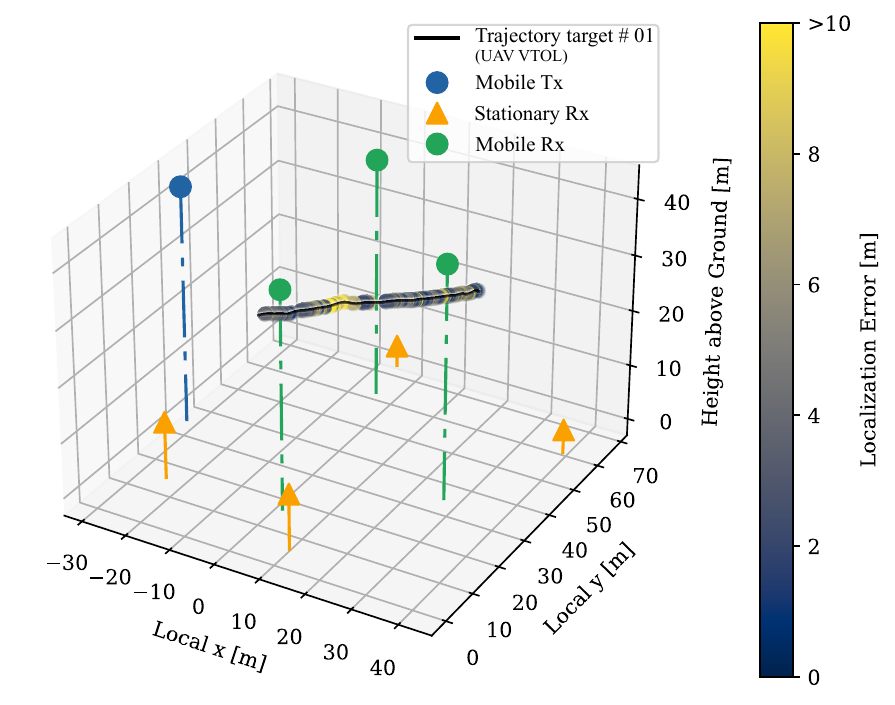}
    \includegraphics[width=0.69\linewidth]{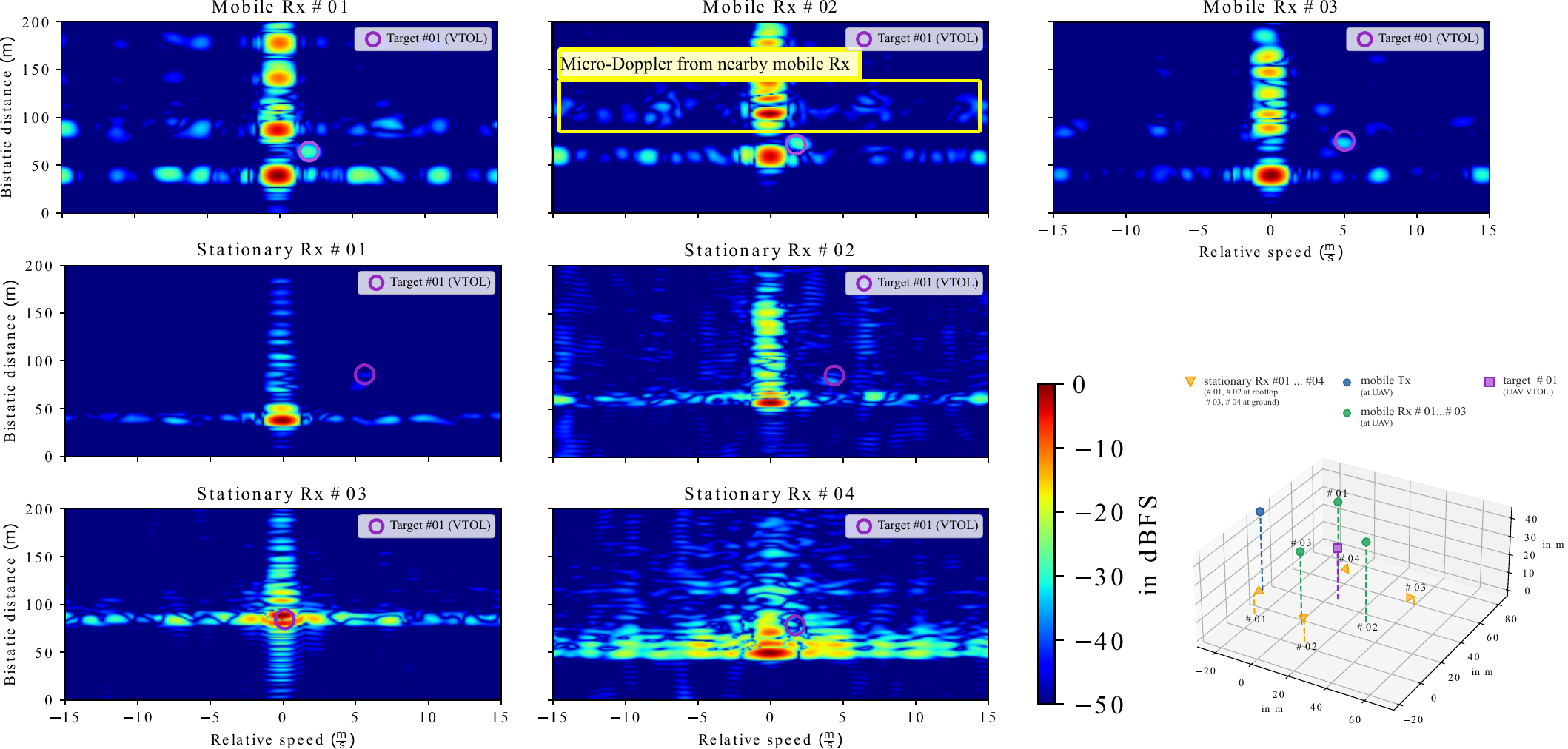}
    \caption{Mobile sensing via a drone swarm in Scenario (a.1). \textit{On the left side:} Localization error between the estimated target positions, derived using least-squares with estimated target delays of all receiver nodes and sensor node position data, and the actual target position obtained from the ground truth unit. \textit{On the right side:} Exemplary delay-Doppler maps with micro-Doppler signatures from the flying sensor node \glspl{uav}.}
    \label{fig:results_a1}
\end{figure*}

To showcase the potential of the data set for swarm-based sensing in \gls{isac} networks, Fig.~\ref{fig:results_a1} illustrates the detection of a large UAV using its reflectivity in a simple line-of-\mbox{sight (LoS)} situation in combination with a basic delay-based radar algorithm.
This setup allows for detailed analysis of Doppler effects and the impact of node mobility on radar sensing accuracy in coordinated \gls{uav} operations, as shown in Fig.~\ref{fig:results_a1}.

\subsection{Infrastructure-based multi-target sensing -- Scenario (d.1)}
\begin{figure*}[h]
    \includegraphics[width=0.3\linewidth]{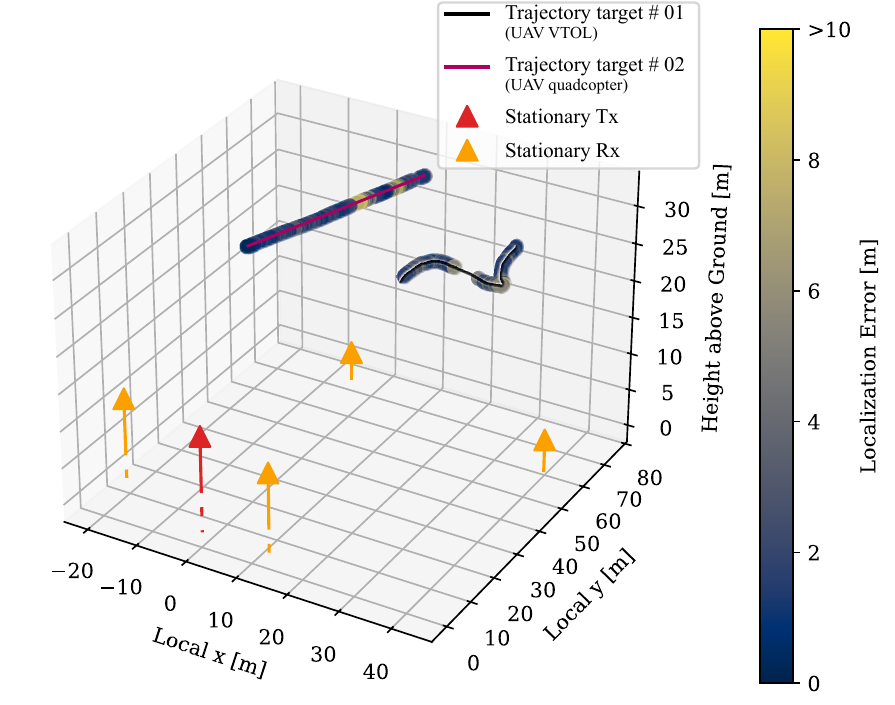}
    \includegraphics[width=0.69\linewidth]{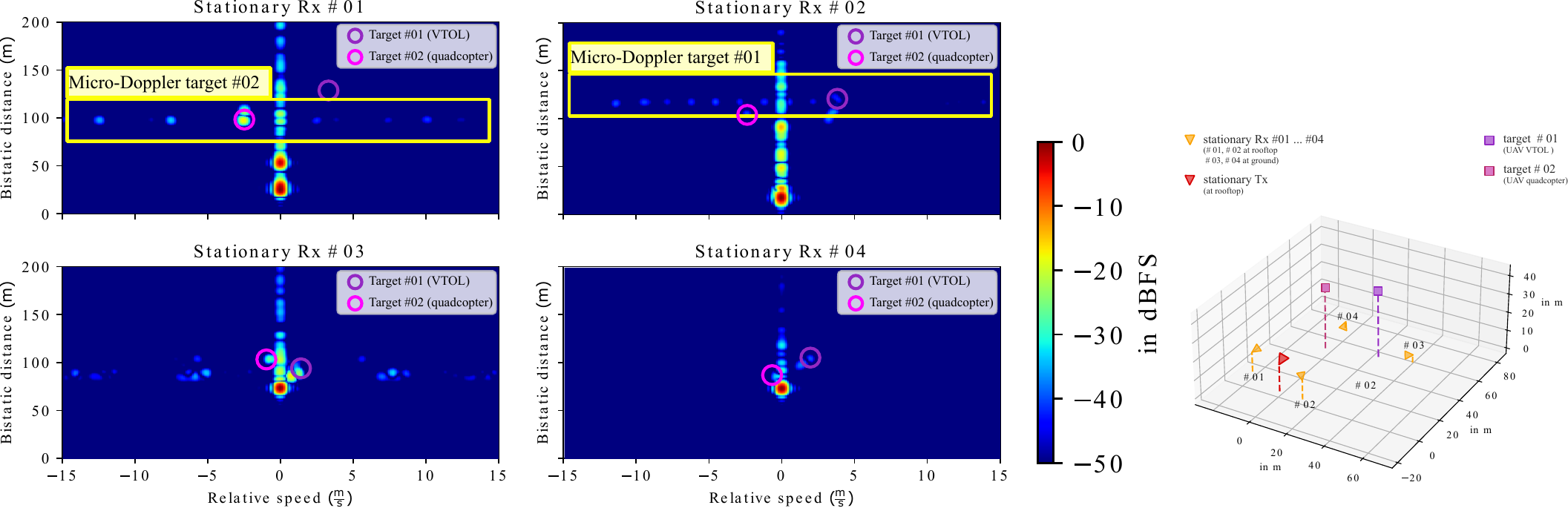}
    \caption{Infrastructure-based multi-target sensing in Scenario (d.1). \textit{On the left side:} Localization error between the estimated positions of the two targets, derived using least-squares with estimated target delays of all receiver nodes and sensor node position data, and the actual target position obtained from the ground truth unit. \textit{On the right side:} Exemplary delay-Doppler maps with micro-Doppler signatures from the flying targets.}
    \label{fig:results_d1}
\end{figure*}

As shown in Fig.~\ref{fig:results_d1}, the data set of scenario (d.1) allows to investigate the detection of multiple flying targets of different sizes using infrastructure-based sensing.

\section{Conclusion}
This paper presents a publicly available real-world data set recorded using a synchronized channel sounding testbed with ground-based and airborne sensor nodes.
The conducted measurement campaign was designed to sense the reflectivity of ground-moving and flying objects, showcasing the potential of collaborating drone swarms for multi-static radar tracking and localization in \gls{a2a} and \gls{a2g} scenarios.
The exemplary results presented in this paper illustrate the capabilities of the data set and highlight the importance of real-word data supplementing simulation-based data to develop robust algorithms for advancing \gls{isac} technologies.

\bibliographystyle{IEEEtran}
\bibliography{paper}

\end{document}